\documentclass[journal,comsoc]{IEEEtran}

\usepackage[T1]{fontenc}

\usepackage{graphicx}
\usepackage{caption}
\usepackage{subcaption}
\usepackage{color}

\IEEEoverridecommandlockouts
\usepackage{mathtools}
\usepackage{amsmath}
\usepackage{algorithmic,algorithm}
\usepackage{cite}

\allowdisplaybreaks

\usepackage{amsmath}
\usepackage{amsthm}

\begin{document}
%
\title{Multi-Robot Association-Path Planning in Millimeter-Wave Industrial Scenarios} 

\author{Cristian~Tatino,~\IEEEmembership{Student Member,~IEEE,}
        Nikolaos~Pappas,~\IEEEmembership{Member,~IEEE,}
        and~Di~Yuan,~\IEEEmembership{Senior Member,~IEEE}
\thanks{This work was supported in part by ELLIIT and by the European Union's Horizon 2020 research and innovation programme under the Marie Sklodowska-Curie grant agreement No. 643002 (ACT5G).

Cristian~Tatino, Nikolaos~Pappas, and Di~Yuan are with Department of Science and Technology (ITN), Link\"{o}ping University,
Sweden (Email: cristian.tatino@liu.se, nikolaos.pappas@liu.se, di.yuan@liu.se)}
}%

\maketitle

\begin{abstract}
The massive exploitation of robots for industry 4.0 needs advanced wireless solutions that replace less flexible and more costly wired networks. In this regard, millimeter-waves (mm-waves) can provide high data rates, but they are characterized by a spotty coverage requiring dense radio deployments. In such scenarios, coverage holes and numerous handovers may decrease the communication throughput and reliability. In contrast to conventional multi-robot path planning (MPP), we define a type of multi-robot association-path planning (MAPP) problems aiming to jointly optimize the robots' paths and the robots-access points (APs) associations. In MAPP, we focus on minimizing the path lengths as well as the number of handovers while sustaining connectivity. We propose an algorithm that can solve MAPP in polynomial time and it is able to numerically approach the global optimum. We show that the proposed solution is able to guarantee network connectivity and to dramatically reduce the number of handovers in comparison to minimizing only the path lengths.
\end{abstract}




%
\IEEEpeerreviewmaketitle

\section{Introduction}
\label{sec:Intro}
The digital transformation of the manufacturing processes that characterizes the fourth industrial revolution (industry 4.0) requires new networking solutions. In this regard, wireless technologies reduce the cost for cable installation and maintenance and they enable the deployment of capillary sensor networks and moving robots for a full industrial automation. Moreover, the increasing throughput demand of new industrial applications, e.g., remote controlling, assembly, and surveillance, makes the millimeter-wave (mm-wave) frequency range (30-300 GHz) an attractive solution~\cite{MMInd,MMInd2}. However, blockage sensitivity at such high frequencies makes the coverage spottier, requiring dense radio deployments. In such scenarios, a robot moving from a starting position to a destination may be subject to coverage holes and numerous handovers that reduce communication throughput and reliability. Namely, a handover requires an initial access phase, the complexity of which is increased by the use of directional beams that need to be aligned. This procedure is costly in terms of energy and time~\cite{PhaseArray}. Thus, both robot path planning and association between access points (APs) and robots need to be optimized to satisfy throughput, reliability, and latency requirements.

Multi-robot path planning (MPP) problems have been analyzed before~\cite{MPP1,MPP2}. In the past few years, joint robot mobility and communication optimization, e.g., motion-transmission energy minimization, has been attracting an increasing amount of interest~\cite{MPC1,MPC2}. However, to the best of our knowledge, none of the previous studies has considered \textit{multi-robot association-path planning} (MAPP) problems.

In this work, we propose a general formulation for MAPP problems in mm-wave scenarios. MAPP aims to jointly find the paths that the robots traverse to reach the respective destinations and the sequence of APs with which they are associated. More precisely, we focus on the type of MAPP with the goal of i) selecting paths for reaching the destinations in the shortest possible time, ii) minimizing the number of handovers, and iii) avoiding coverage holes, robot collisions, and AP overloading. To solve the MAPP problems, we propose an algorithm that is based on a column generation scheme and can run in polynomial time. The algorithm can dramatically reduce the number of handovers per robot with a slightly increase in the path lengths in comparison to minimizing only the latter.

\begin{figure}[tb]
	\centering
	\includegraphics[width=7cm]{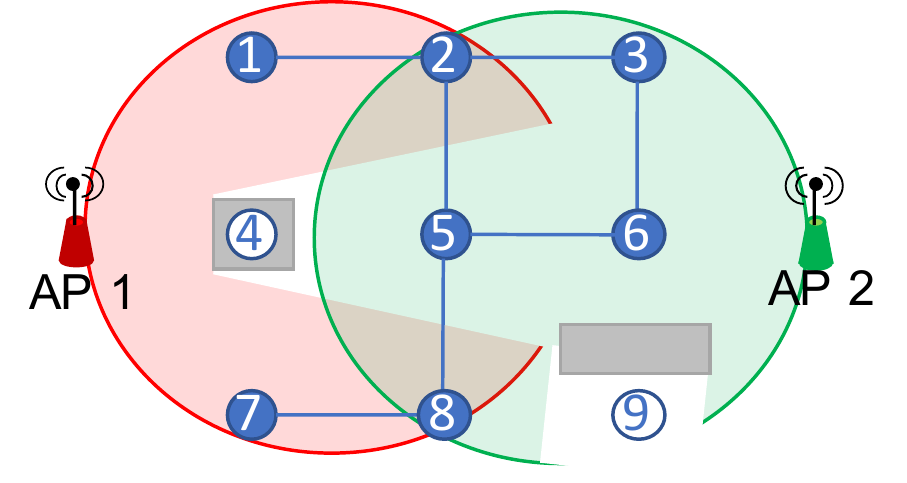}
	\caption[]{A scenario consisting of a 3 x 3 grid forming a graph with 7 vertices and 7 edges covered by two APs. Vertices and edges of positions 4 and 9 are not included in $G$.  Namely, the former position is occupied by an obstacle, whereas the latter position is not covered neither by AP 1 because it is too far, nor AP 2 because of a blockage.}
	\label{fig:Scen}
\end{figure}

\section{System Model and Assumptions}
\label{sec:Ass}
We consider an industrial scenario, e.g., an industrial plant, where a set of $N$ robots need to move from their starting positions $o_{i}$ to their goals $d_{i}$, with $1 \le i \le N$, within a time horizon of duration $T$. The robots can move on an undirected graph $G=(\mathcal{V},\mathcal{E})$ that is covered by a set $\mathcal{A}$ of $A$ APs using mm-wave. More precisely, $\mathcal{V}$ is the set of vertices and $\mathcal{E}$ the set of edges, with cardinality $V$ and $E$, respectively. Each vertex $v \in \mathcal{V}$ represents a physical position with coordinates $(x_{v},y_{v})$. An edge $e=\{v,u\} \in \mathcal{E}$, with $v,u \in \mathcal{V}$, represents a segment between the points $(x_{v},y_{v})$ and $(x_{u},y_{u})$. Some positions may be occupied by 3-dimensional (3D) obstacles with certain sizes. As shown in Fig.~\ref{fig:Scen}, vertices and edges of $G$ are defined only for positions that are free from obstacles and covered by at least one AP. This guarantees network connectivity. For this reason, we assume a radio map that for each vertex and edge of $G$ provides the APs that cover them.

The radio map can be either obtained by measurements, which are easily collected for highly controlled scenarios like industrial ones, or by computing the signal-to-noise ratio (SNR) at each position. By using the second method, a robot is covered by AP $a$ if and only if (iff) the SNR at the receiver is higher than a threshold $\gamma$, i.e., $\mbox{SNR}_{a}(x,y)\ge \gamma$. In order to obtain $\mbox{SNR}_{a}(x,y)$, we first compute whether a robot at the position $(x,y)$ is in line-of-sight (LOS) or non line-of-sight (NLOS) with the $a$-th AP. This depends only on the positions and the heights of the robots, the obstacles, and the APs, which are assumed to be known. Both robots and APs use directional transmissions with a certain beamwidth. When this is narrow enough, we can assume that the interference among robots becomes negligible~\cite{Inter}. Moreover, we assume that the receiving antenna is located at the center on the top of the robots. These have the same height, thus, they can not obstruct each other's LOS with an AP. In Section~\ref{sec:Res}, we provide more details of the channel model and its parameters.

We consider slotted time, $t=0,...,T$ and in each timeslot, a robot may either stay at the current vertex or move to an adjacent one. We assume that the robots move at a constant speed and take one timeslot to traverse an edge. Moreover, in any timeslot, an edge or vertex can be traversed or occupied by at most one robot. At each position, a robot is associated with one AP. While traversing an edge or remaining at a vertex, a robot may face a handover, maximum one per timeslot. A handover to a new AP is needed when: i) load balancing among the APs is necessary, or ii) the robot exits the coverage area of the currently associated AP. The latter event occurs mainly because either the distance between the robot and the AP becomes too long or because one or multiple obstacles block the signal~\cite{BLO}. For this reason, the selection of both the paths and the AP association must be optimized.

\section{Problem Formulation}
\label{sec:Prob}
In this section, we first provide a formulation for MAPP problems as an integer linear program (ILP) that aims to minimize the total robot path cost, while avoiding robot collisions, and AP overloading. As explained in Section~\ref{sec:Ass}, the connectivity at each position with at least one AP is guaranteed by the radio map and the definition of $G$. The path cost can be defined in order to minimize several objective functions. In this work, we focus on a particular instance of MAPP of which the priority is to minimize the number of handovers (MAPP-HP). Since multiple paths can have the same number of handovers, MAPP-HP selects the paths with the shortest traversal times among those of minimum number of handovers.

We consider a path-based formulation as in~\cite{ColGen}, where a path of a robot is fully described by an ordered set of tuples. Each tuple consists of an edge, the timeslot when the robot enters the edge, and the associated AP, e.g., $(\{v,u\},t,a)$. For each robot $i$, we consider the sets $\mathcal{S}_{i}$ of all the possible paths that connect the source $o_{i}$ and the destination $d_{i}$. For each path, we define a cost $c_{is}$ and a binary variable $x_{is}$ that is equal to $1$ if the $i$-th robot uses path $s \in \mathcal{S}_{i}$ and $0$ otherwise. Moreover, we define the following binary parameters:
\begin{itemize}
\item $b_{iets}$ is equal to $1$ if path $s$ of the $i$-th robot enters edge $e$ at time $t$, and $0$ otherwise,
\item $g_{ivts}$ is equal to $1$ if path $s$ of the $i$-th robot stays at vertex $v$ at time $t$, and $0$ otherwise,
\item $l_{iats}$ is equal to $1$ if along path $s$, the $i$-th robot is associated with AP $a$ at time $t$, and $0$ otherwise.
\end{itemize}
Then, we can write the following ILP:
\begin{subequations}
\begin{align}
        MAPP:&\min_{x_{is}} \sum_{i=1}^{N}\sum_{s \in \mathcal{S}_{i}}c_{is} x_{is}\label{opt}\\ 
        \text{s.t.}&\sum_{s \in \mathcal{S}_{i}}x_{is}= 1, \label{SD_con} \forall i=1,...,N,\\
        		& \sum_{i=1}^{N}\sum_{s \in \mathcal{S}_{i}}b_{iets}x_{is} \le 1, \forall e\in\mathcal{E}, t=1,...,T, \label{Con_Edges}\\
		&\sum_{i=1}^{N}\sum_{s \in \mathcal{S}_{i}}g_{ivts}x_{is} \le 1, \forall v\in\mathcal{V}, t=1,...,T, \label{Con_Vert}\\
		&\sum_{i=1}^{N}\sum_{s \in \mathcal{S}_{i}}l_{iats}x_{is} \le m, \forall a\in\mathcal{A}, t=1,...,T,\label{Con_Load}\\
		 &x_{is}\in\{0,1\} \label{type} \quad \forall i=1,...,N, s \in\mathcal{S}_{i}.
\end{align}
\end{subequations}
The objective function~\eqref{opt}, represents the sum of the robots' path costs, which, for MAPP-HP, are defined in Section~\ref{sec:Graph}. Constraint~\eqref{Con_Edges} prevents multiple robots from traversing the same edge $e$ in the same timeslot, whereas, constraint~\eqref{Con_Vert} allows at most one robot per timeslot to stay at a vertex $v$. Finally,~\eqref{Con_Load} limits the number of robots that are simultaneously associated to an AP $a$ to be at most $m$.

MAPP-HP is NP-hard. Namely, MPP for traversal time minimization (MTATMPP) has been proven to be NP-hard in~\cite{NPH}. Since MTATMPP is equivalent to MAPP-HP with only one AP, we have that MAPP-HP is NP-hard. Moreover, the cardinality of $\mathcal{S}_{i}$ grows exponentially with $T$, the number of edges, and the number of APs. However, most of the paths are not relevant for constructing the optimal solution. Therefore, to solve MAPP problems, we consider an algorithm based on a column generation scheme that is presented in the next section.

\section{Algorithm}
\label{sec:Alg}
In this section, we present a \textit{column generation} based algorithm to deal with the exponential growth of paths, and thereby solving MAPP problems. The basic idea of column generation is to solve a linear programming problem (LP) for MAPP with a restricted set of variables (paths) and then add paths that may improve the solution. Before applying column generation, we first expand $G$ to a directed graph $G'$, whose edges and vertices include association and handover information. This allows us to use shortest path algorithms to both find an initial solution and generate new paths. Then, we construct a continuous relaxation of the restricted MAPP, with restricted set of paths $\mathcal{\hat{S}}_{i}$, called \textit{master problem}. The initial solution of the master problem may be far from optimality. For this reason, we expand the paths of the master problem by adding new ones that can improve the objective function. Then, we find an integer solution by reducing the generated paths to only one per robot, as will be explained in Section~\ref{sec:Integ}. The resulting algorithm is called \textit{Path Generation with Cooperative Pruning} (\textit{PGCP}). This is shown in Algorithm~\ref{alg:Algo} and described in the following sections.

\begin{figure}[tb]
	\centering
	\includegraphics[width=7cm]{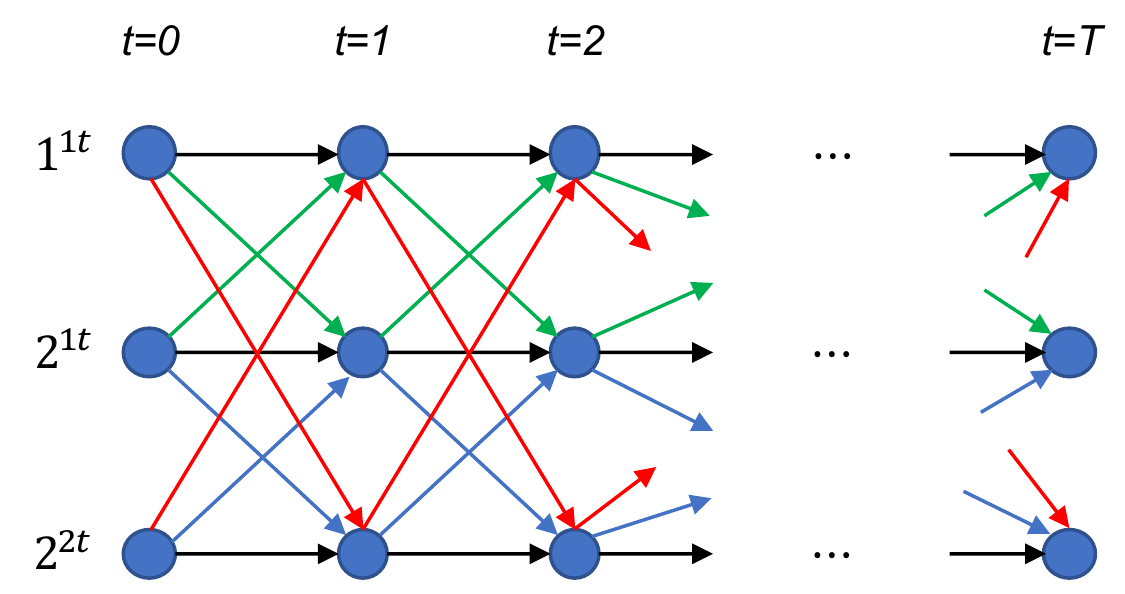}
	\caption[]{Time-coverage expansion of vertices 1 and 2 of Fig.~\ref{fig:Scen}. Vertex 1 is covered by AP 1, whereas vertex 2 is covered by both the APs. Edge $\{1,2\}$ in $G$ corresponds to either the red or the green edges of $G'$, depending on whether the robot is performing an handover or not, respectively. Black and blue edges represent waiting times at a vertex, with the latter including a handover.}
	\label{fig:ExpG}
\end{figure}

\subsection{Time-Coverage Expanded Graph}
\label{sec:Graph}
In this section, we expand the graph $G$ to a directed time-coverage expanded graph $G'=(\mathcal{V'},\mathcal{E'})$ with $\mathcal{V'}$ and $\mathcal{E'}$ being the expanded vertex and edge sets, respectively. For each vertex $v \in \mathcal{V}$, AP $a$ that covers $v$, and timeslot $t=0,...,T$, we create a copy $v^{at} \in \mathcal{V'}$. Then, for each edge $e=\{v,u\} \in \mathcal{E}$ and timeslot $t$, we define an edge $e'=\{v^{at},u^{bt+1}\} \in \mathcal{E'}$ for any two APs $a,b \in \mathcal{A}$ that cover vertices $v$ and $u$, respectively. Moreover, to represent the waiting time of a robot at a vertex $v$, for any two APs $a,b \in \mathcal{A}$ that cover $v$, we add an edge $e'=\{v^{at},v^{bt+1}\} \in \mathcal{E'}$ between any two consecutive timeslots. An example of expanded graph is given in Fig~\ref{fig:ExpG}. For MAPP-HP, the cost $c_{e'}$ for an edge $e'=\{v^{at},u^{bt+1}\}$ is defined as follows:
\begin{align}
c_{e'} = \begin{cases} c_{h}+c_{t} &\mbox{if} \; a\neq b,\\
c_{t} &\mbox{otherwise},\\
\end{cases}
\label{eq:cost}
\end{align}
where, $c_{h}$ is the handover cost, and $c_{t}$ represents the traversal time of the edge that is set to $1$ timeslot\footnote{This work can be generalized to the case of edges with different traversal times by modifying the expanded graph.}. Since in MAPP-HP, we penalize the handovers more than the traversal time, we set $c_{t} \ll c_{h}=T$, where $T$ is the time horizon that is the maximum possible traversal time for a path. The cardinalities of $\mathcal{V'}$ and $\mathcal{E'}$ are $VA(T+1)$ and $(2E+V)TA^{2}$, respectively.

\subsection{Path Generation}
\label{sec:Path}
Given $G'$, we find an initial solution by using the \textit{cooperative A*} algorithm~\cite{CoopA}. This provides one path for each robot that are added to the restricted sets $\mathcal{\hat{S}}_{i}$ of the master problem and converted to constraints~\eqref{Con_Edges},~\eqref{Con_Vert}, and~\eqref{Con_Load}. However, cooperative A* may fail to find a feasible solution. Thus, we add an artificial path for each robot's source-destination pair with a cost much higher than any real path.

Starting from the initial solution, we solve the master problem and add new paths that can improve the current solution. Namely, for each robot, we find the path with the minimum reduced cost by solving the following \textit{pricing problem}:
\begin{align*} 
\min_{s \in \mathcal{\hat{S}}_{i}} \; & c_{is}-\phi_{i}-\sum_{e \in \mathcal{E}}\sum_{t=0}^{T}\pi_{et}b_{iets}-\sum_{v \in \mathcal{V}}\sum_{t=0}^{T}\gamma_{vt}g_{ivts}\\
&-\sum_{a \in \mathcal{A}}\sum_{t=0}^{T}\lambda_{at}l_{iats}\stepcounter{equation}\tag{\theequation}\label{eq:RC},
\end{align*}
where, the objective~\eqref{eq:RC} is to find the path of minimum reduced cost of robot $i$. Then, $s$ is added to $\mathcal{\hat{S}}_{i}$ only when its reduced cost is negative. The path generation concludes when there are no more paths having negative reduced costs among those that are not included in subsets $\mathcal{\hat{S}}_{i}$. The term $c_{is}$ of~\eqref{eq:RC}  is the sum of the edges' cost $c_{e'}$ that belong to path $s$. The terms $\phi_{i}$, $\pi_{et}$, $\gamma_{vt}$, and $\lambda_{at}$ are the dual variables associated with constraints~\eqref{SD_con}, \eqref{Con_Edges}, \eqref{Con_Vert}, and~\eqref{Con_Load}, respectively. Note that $-\pi_{et}$ contributes to the reduced cost iff $b_{iets}$ is equal to $1$. This occurs when the $i$-th robot enters edge $e$ at time $t$. Namely, for each edge $e=\{v,u\} \in \mathcal{E}$ of the original graph $G$, we can add $-\pi_{et}$ to cost $c_{e'}$ of the corresponding edges on the expanded graph $G'$, i.e., $e'=\{v^{at},u^{bt+1}\} \in \mathcal{E'}, \; \forall \; a,b \in \mathcal{A}$. The same reasoning can be applied for $g_{ivts}$ and $l_{iats}$ and we add $-\gamma_{vt}$ to the cost of the edges that enter vertex $v$ at time $t$. Moreover, we add $-\lambda_{at}$ to the cost of those edges that, at time $t$, enter a vertex that is covered by AP $a$.

Thus, we can minimize~\eqref{eq:RC}, by finding the shortest path from $o_{i}$ to $d_{i}$ on the expanded graph $G'$ with the edge costs modified by the dual variables $\pi_{et}$, $\gamma_{vt}$, and $\lambda_{at}$. This problem can be solved in polynomial time. More specifically, in this work, we use the A* algorithm.

\subsection{Finding an Integer Solution: Cooperative Pruning}
\label{sec:Integ}
When the path generation concludes, we can not guarantee that the solution of the master problem is integer. To find an integer solution, we proceed as follows. For each robot $i$, we construct a mixed-integer linear program (MILP) from the master problem by setting the variables $x_{is}$ to be binary. The variables corresponding to the other robots, i.e., $x_{js}$, with $j \neq i$, remain continuous. Since constraints~\eqref{SD_con}, the solution of MILP has exactly one variable $x_{is}$ that is equal to one. If this does not correspond to the artificial path, we delete all the other paths and the corresponding variables $x_{ip}$ with $p \neq s$ from the master problem. Otherwise, if the artificial path variable is equal to 1, we continue generating paths, as done in Section~\ref{sec:Path} until none of them has negative reduced cost. This repeats until all robots have one path selected. In case any robot uses the artificial path and no more path is generated, the algorithm declares infeasibility.

\subsection{Algorithm Complexity}
\label{sec:Comp}
In this section, we conclude that \textit{PGCP} can run in polynomial time. We first note that the pricing problem represents a separation problem for the dual of the master~\cite{Wolsey}. Now, we can use Theorem 3.3 on p. 163 in~\cite{Wolsey}. Namely, an LP is solvable in polynomial time iff the separation problem is solvable in polynomial time. In \textit{PGCP}, the separation problem is a shortest path problem solved by algorithm A* that is polynomial. This result applies to both the path generation part and the column generation for finding an integer solution. Moreover, solving the MILP for robot $i$ (in Line 16) is equivalent to solving several LPs, one for each path $s \in \mathcal{S}_{i}$ with $x_{is}=1$, and consider the best solution. 

\begin{algorithm}
\algsetup{linenosize=\tiny}
\caption{\textit{PGCP}}
\label{alg:Algo}
\begin{algorithmic}[1]
  \scriptsize
  \STATE{Construct expanded graph $G'$}
  \item[\textbf{\textit{Initial solution}}:]
  \STATE{Find initial paths with \textit{Cooperative A*} and add them to $\mathcal{\hat{S}}_{i}$}
  \STATE{Add artificial paths to $\mathcal{\hat{S}}_{i}$}
  \item[\textbf{\textit{Path Generation}}:]
  \REPEAT {
 	 \STATE{Construct the master problem from restricted ILP with paths in $\mathcal{\hat{S}}_{i}$} 
  	 \STATE{Solve the master problem and add $\pi_{et}$, $\gamma_{vt}$, and $\lambda_{at}$ to edge costs $c_{e'}$}
  	\FOR{each robot $i \in \mathcal{R}$}
		\STATE{Compute shortest path on $G'$ with \textit{A* algorithm}}
		\IF{path cost $<0$}
			\STATE{Add the path to $\mathcal{\hat{S}}_{i}$}
		\ENDIF 
  	\ENDFOR
	 } \UNTIL{no new paths are added to $\mathcal{\hat{S}}_{i}, \forall i \in \mathcal{R}$}
  \item[\textbf{\textit{Cooperative Pruning}}:]
  \REPEAT {
    	\FOR{each robot $i \in \mathcal{R}$}
		\STATE{Construct MILP from the master problem with $x_{is}=\{0,1\}$}
		\STATE{Solve MILP}
		\IF{MILP solution is feasible} 
			\STATE{$\forall s \in \mathcal{\hat{S}}_{i}: x_{is}=0$ delete path $s$ from $\mathcal{\hat{S}}_{i}$}
		\ELSE
			\STATE{Repeat Steps 8-11}
		\ENDIF 
  	\ENDFOR	  
  } \UNTIL{no new paths are added to $\mathcal{\hat{S}}_{i}, \forall i \in \mathcal{R}$}
\end{algorithmic}
\end{algorithm}

\section{Numerical Results}
\label{sec:Res}
In this section, we provide a numerical evaluation of \textit{PGCP} for solving MAPP-HP. The result of \textit{PGCP} is compared with the initial solution obtained by Cooperative A*. Moreover, we show the results of applying \textit{PGCP} to other two MAPP objectives: MAPP with traversal time priority (MAPP-TP) and MAPP with maximum SNR criteria (MAPP-SNR). The former considers an opposite criterion to MAPP-HP. Namely, MAPP-TP jointly minimizes path traversal time and the number of handovers with the former having priority over the latter, i.e., $c_{h} \ll c_{t}$. MAPP-SNR has the same formulation of MAPP and minimizes only the total traversal time, i.e., $c_{h}=0$, while selecting the AP with the maximum SNR at each position.

For our simulations we consider a grid of 20 x 20 vertices covering a square-shaped indoor scenario with a side length of $60$\,m. Obstacles, with a height of  $2$\,m, are randomly dropped and occupy almost $30 \%$ of the vertices. There are $4$ APs, operating at different frequency channels in the $60$~GHz band. The APs are equally distributed and placed at a height of $5$\,m, whereas, we set the antenna height at the robot equal to $0.5$\,m. To compute the coverage and the radio map, we use the 3GPP model for indoor scenarios~\cite{3GPP} for computing the SNR. The model considers the distance between the AP and the robot, whether they are in LOS or NLOS, and several other parameters. More precisely, we set the transmit and the noise powers  to $24$\,dBm and $-80$\,dBm, respectively. The antenna gain at the APs and at the robots are $g_{a}=15$\,dB and $g_{r}=1$\,dB, respectively. The SNR threshold is equal to $\gamma = 10$\,dB. Unless specified otherwise, the time horizon is $T=60$ timeslots. With these parameters, the resulting time-coverage expanded graph has an average of $50000$ vertices and $600000$ edges.

\begin{figure}[tb]
	\centering
	\includegraphics[width=7cm]{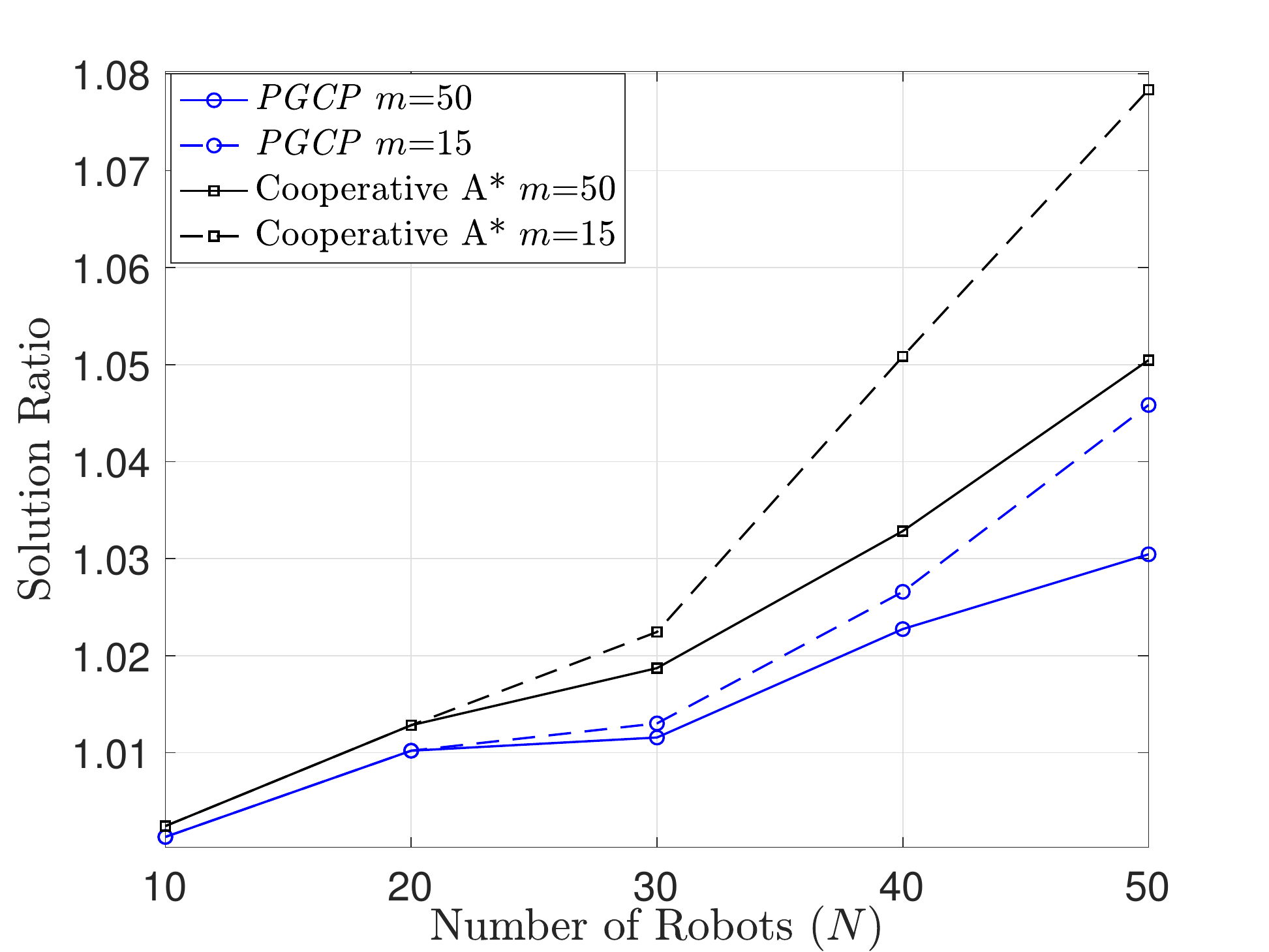}
	\caption[]{Solution ratio for \textit{PGCP} and cooperative A* for MAPP-HP with varying $N$ and $m$.}
	\label{fig:Gap}
\end{figure}
\begin{table}[]
\centering
\caption{Success Rate of \textit{PGCP} and Cooperative A* for MAPP-HP.}
\label{undefined}
\begin{tabular}{c|c|c|c|c|}
\cline{2-5}
 & \multicolumn{2}{c|}{$m=15$} & \multicolumn{2}{c|}{$m=50$} \\ \cline{2-5} 
 & Cooperative A* & \textit{PGCP} & Cooperative A* & \textit{PGCP} \\ \hline
\multicolumn{1}{|c|}{$T=60$} & $67$ \% & $91$ \% & $83$ \% & $95$ \% \\ \hline
\multicolumn{1}{|c|}{$T=90$} & $84$ \% & $96$ \% & $89$ \% & $99$ \% \\ \hline
\end{tabular}
\end{table}

In Fig.~\ref{fig:Gap}, we show the solution ratio for both \textit{PGCP} and Cooperative A* (initial solution) for MAPP-HP with respect to the number of robots ($N$) and the maximum number of robots associated per AP ($m$). The solution ratio of \textit{PGCP} is defined as the ratio between the total path costs obtained by \textit{PGCP} and the solution of the master problem when the path generation ends. The latter is guaranteed to be a lower bound to the optimal solution of MAPP-HP. The same ratio is defined for Cooperative A*. We can observe that \textit{PGCP} is able to approach the global optimum and the solution ratio is an increasing function of $N$, whereas, it decreases with increasing values of $m$. This is more clear for $m \ll N$. Moreover, when $N$ is small with respect to the dimension of the graph, \textit{PGCP} and cooperative A* have similar solution ratios. Namely, the robots are not in conflict with each other in the choice of the paths and APs. 

Note that, the results shown in the figures of this section, are based on cases for which feasible solutions are found by both algorithms. Both \textit{PGCP} and cooperative A* can terminate with infeasible solutions. However, as we show in Table I, \textit{PGCP} can provide a higher percentage of feasible solutions (success rate) than cooperative A*. We can observe that higher values of $m$ and $T$ lead to a higher number of feasible paths that increases the success rate. However, higher values of $T$ lead to larger time-coverage expanded graphs that increase the computational time. For the analyzed scenario, this is approximately $30$\,secs for \textit{PGCP} with $N=50$ and $T=60$, with a laptop with 8 GB of RAM and a 7th generation, Intel Core i7 processor. This value can be further improved by parallelizing the implementation of \textit{PGCP}.
\begin{figure}[tb]
	\centering
	\includegraphics[width=7cm]{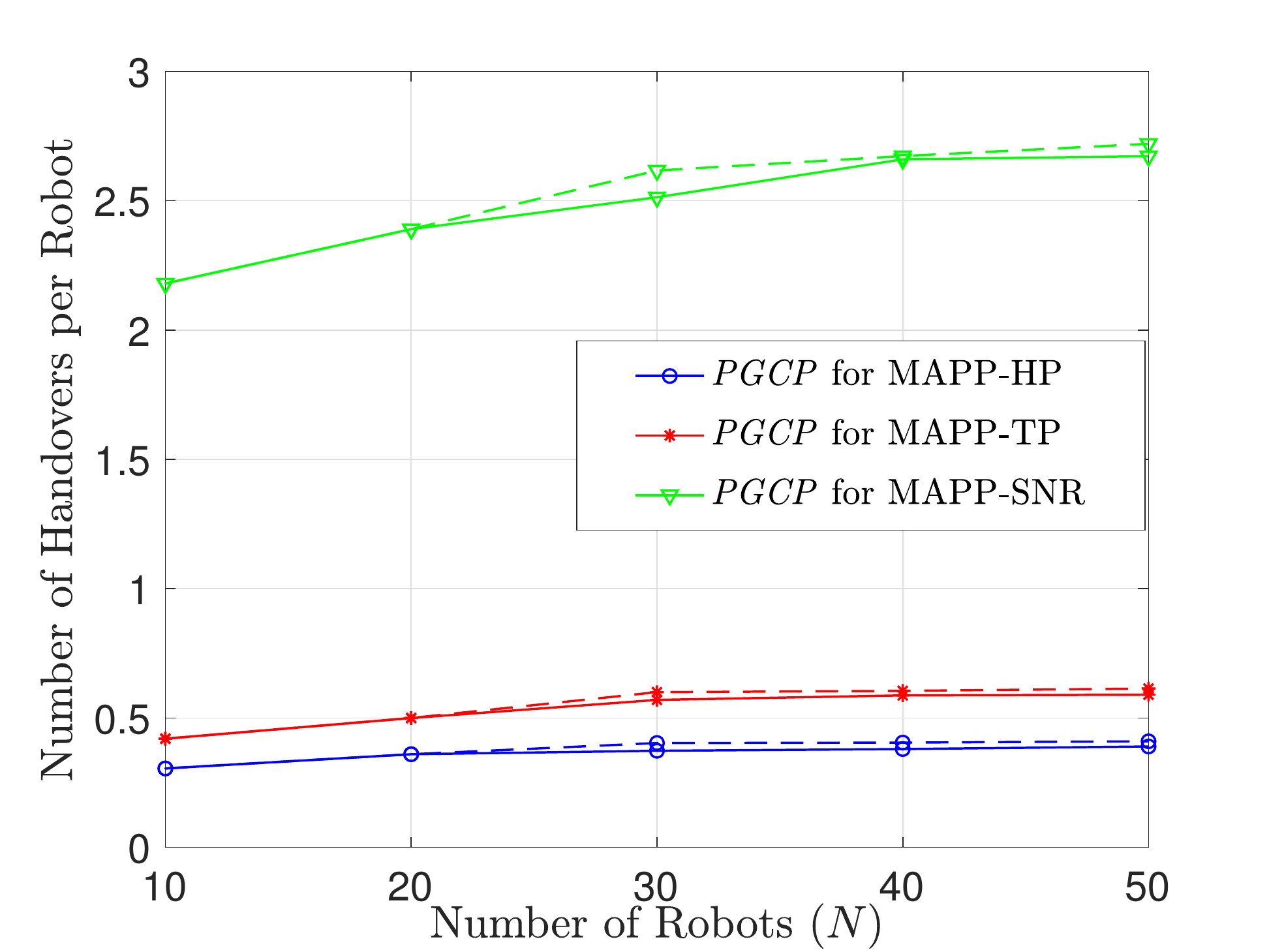}
	\caption[]{Average number of handovers per robot by applying \textit{PGCP} to MAPP-HP, MAPP-TP, and MAPP-SNR. We vary $N$ and show the results for $m=50$ and $m=15$ with solid and dashed lines, respectively.}
	\label{fig:Hand}
\end{figure}

\begin{figure}[tb]
	\centering
	\includegraphics[width=7cm]{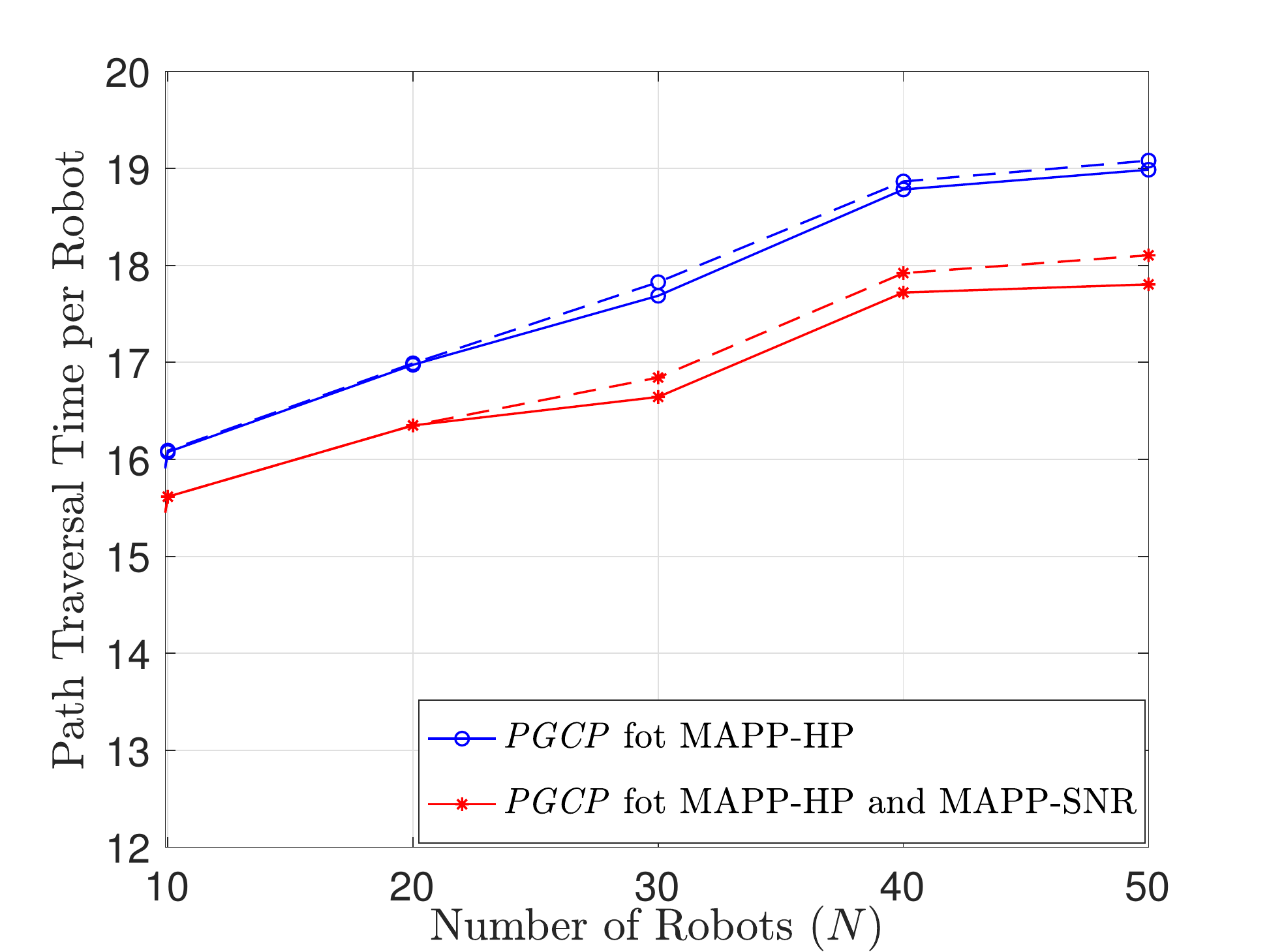}
	\caption[]{Average path traversal time per robot by applying \textit{PGCP} to MAPP-HP and MAPP-TP with varying $N$ and $m$. We show the results for $m=50$ and $m=15$ with solid and dashed lines, respectively. Note that MAPP-SNR presents the same average path traversal time of MAPP-TP.}
	\label{fig:Len}
\end{figure}

Now, we separately analyze the two components of the path cost, namely handover and traversal time. In Fig.~\ref{fig:Hand} and Fig.~\ref{fig:Len}, we show the average number of handovers and the average path traversal time per robot by applying \textit{PGCP} to MAPP-HP, MAPP-TP and MAPP-SNR. In Fig.~\ref{fig:Hand}, as expected, the solution of MAPP-SNR presents the highest number of handovers, whereas, an optimized selection of APs is able to dramatically reduce the handovers per robot. In this regard, MAPP-HP reduces the number of handovers by $50$\% with respect to MAPP-TP. As shown in Fig.~\ref{fig:Len}, MAPP-TP provides the shortest traversal time per robot, which coincide with that of MAPP-SNR. However, optimizing handovers by MAPP-HP results in an increase of only the $5$\% with respect to MAPP-TP that prioritizes the traversal time. We can observe that both in Fig.~\ref{fig:Hand} and Fig.~\ref{fig:Len}, for all the presented MAPP objectives, the handovers and the traversal time per robot increase either when $N$ increases or $m$ decreases. More precisely, as introduced also for Fig.~\ref{fig:Gap}, when $N$ is large with respect to the value of $m$ and to the graph dimension, it is more likely that the robots' paths diverge from the optimal ones in order to avoid collisions or AP overloading. This results in longer traversal times and higher number of handovers.



\section{Conclusion}
\label{sec:Conc}
In this work, we have proposed a novel type of multi-robot association-path planning (MAPP) problems. In contrast to conventional robot path planning, MAPP takes into consideration the radio coverage and aims to jointly optimize robots' paths and robot-AP associations. The optimization of MAPP problems can be fundamental to satisfy throughput and reliability requirements in mm-wave industrial scenarios. 

We proposed an algorithm (\textit{PGCP}) based on a column generation scheme for solving several MAPP objectives. These aim to jointly minimize the number of handovers and the path traversal time. We have shown that \textit{PGCP} can solve MAPP in polynomial time and it is able to numerically approach the global optimum. Moreover, \textit{PGCP} is able to improve the initial solution (provided by cooperative A*) while guaranteeing higher success rate. When handovers are prioritized over the path traversal time, \textit{PGCP} can reduce the handovers by $50\%$ with respect to problems that aim first to minimize the path lengths and then optimize the robot-AP association. The gain in terms of handovers is even higher with respect to solutions without an optimized robot-AP association policy.

\bibliography{ref}
\bibliographystyle{IEEEtran}

\end{document}